\documentclass[revtex4-2, pdflatex,sn-mathphys-num]{sn-jnl}

\usepackage{graphicx}%
\usepackage{amsmath,amssymb,amsfonts}%
\usepackage{siunitx}
\usepackage{braket}
\DeclareSIUnit{\bar}{bar}

\begin{document}

\title[Strong coupling of a superconducting flux qubit to single bismuth donors]{Strong coupling of a superconducting flux qubit to single bismuth donors}

\author[1]{\fnm{T.} \sur{Chang}}
\author[1]{\fnm{I.} \sur{Holzman}}
\author[2]{\fnm{S. Q.} \sur{Lim}}
\author[2]{\fnm{D.} \sur{Holmes}}
\author[3]{\fnm{B. C.} \sur{Johnson}}
\author[2]{\fnm{D. N.} \sur{Jamieson}}
\author*[1]{\fnm{M.} \sur{Stern}}\email{michael.stern@biu.ac.il}

\affil[1]{
 \orgdiv{Quantum Nanoelectronics Laboratory},
 \orgname{Department of Physics \& Bar-Ilan Institute of Nanotechnology and Advanced Materials (BINA)},
 \orgaddress{\city{Ramat-Gan}, \postcode{5290002}, \country{Israel}}}

\affil[2]{\orgdiv{ARC Centre for Quantum Computation and Communication Technology
(CQC-T) \& School of Physics}, \orgname{University of Melbourne},
 \orgaddress{ \city{Parkville}, \postcode{3010}, \state{VIC}, \country{Australia}}}
\affil[3]{\orgdiv{School of Science}, \orgname{RMIT University}, \orgaddress{\city{Melbourne}, \postcode{3000}, \state{Victoria}, \country{Australia}}}


\abstract{
The realization of a quantum computer represents a tremendous scientific
and technological challenge due to the extreme fragility of quantum
information. The physical support of information, namely the quantum
bit or qubit, must at the same time be strongly coupled to other qubits
by gates to compute information, and well decoupled from its environment
to keep its quantum behavior. An interesting physical system for realizing
such qubits are magnetic impurities in semiconductors, such as bismuth donors
in silicon. Indeed, spins associated to bismuth donors can reach an extremely
long coherence time - of the order of seconds. Yet it is extremely
difficult to establish and control efficient gates between these spins. Here we demonstrate a protocol where single bismuth donors can coherently
transfer their quantum information to a superconducting flux qubit, which
acts as a mediator or quantum bus. This superconducting device allows
to connect distant spins on-demand with little impact on their coherent
behavior.}

\maketitle

\newcommand{\hamil}{\mathcal{H}}
\newcommand{\ifluc}{\delta I}
\newcommand{\silicon}{\text{Si}}
\newcommand{\siox}{\text{SiO}_{2}}
\newcommand{\sipure}{^{28}\text{Si}}
\newcommand{\sibad}{^{29}\text{Si}}
\newcommand{\sibi}{\text{Si:Bi}}
\newcommand{\vbo}{\vec{B}}
\newcommand{\vso}{\vec{S}}
\newcommand{\vio}{\vec{I}}
\newcommand{\im}{\boldsymbol{i}}

Spins in semiconductors are often considered as one of the major candidates
for quantum information processing \citep{RevModPhys.85.961,RevModPhys.95.025003}.
They can be extremely well isolated from their environment by choosing
a proper ultrapure surrounding lattice, that contains only nuclei
with zero spin. Consequently, they can safely store quantum information
with low error rates \citep{Muhonen2014}. The primary obstacle in
developing a spin-based quantum processor is the realization of gates
between distant individual spins \citep{Samkharadze2018, Landig2018, PhysRevX.12.021026}. Indeed, quantum computation requires
in addition to single qubit gates, a two qubit controlled rotation
which rotates the spin of a target qubit if and only if the control
qubit is oriented in a given direction. For impurity donors, one possible
approach consists of directly coupling adjacent spins by electrically
controlling their exchange interaction \citep{Kane1998}. Although
this method can potentially yield coupling strengths on the order
of GHz \citep{He2019}, it requires positioning the donor atoms less
than $\SI{15}{\nano\meter}$ apart. This proximity requirement presents
substantial challenges for the fabrication of the ancillary structures
needed for control and readout of the qubit register.

Indirect coupling strategies involve using a common intermediate element
such as a superconducting resonator to couple distant spins at long
distances \citep{RevModPhys.85.623,Kurizki2015,Morello2020}. The
magnetic coupling strength of a single electron spin to a superconducting
resonator is however extremely small. It can be estimated from the
amplitude of the magnetic field fluctuations, $\delta B$, generated
by the circuit. Assuming an infinitely thin wire, the field fluctuations
are given by Biot and Savart law as $\delta B=\tfrac{\mu_{0}\delta I}{2\pi r}$
where $\delta I=\omega_{r}\sqrt{\hbar/(2Z_{0})}$ are the current
fluctuations in the wire, $r$ is the distance from the wire to the
spin, $Z_{0}$ the characteristic impedance of the resonator and $\omega_{r}$
its resonance frequency \citep{PhysRevA.95.022306}. Assuming the
electron behaves as a point-like particle and the magnetic field $\delta B$
is oriented in a direction perpendicular to the spin axis, the magnetic
coupling is given by $\hamil_{c}=\frac{\hbar}{2}\gamma_{e}\delta B\,\sigma_{s}^{x}\left(a+a^{\dagger}\right)$
where $\gamma_{e}/2\pi=\SI{27.997}{\giga\hertz\per\tesla}$ is the
electron gyro-magnetic ratio, $\sigma_{s}^{x}$ is a Pauli operator
acting on the spin degree of freedom and $a$ and $a^{+}$ are respectively
the annihilation and creation operators of an excitation (photon)
in the resonator. We thus obtain the coupling constant,
\begin{equation}
\frac{g}{2\pi}=2.8\frac{\ifluc\left[\si{\nano\ampere}\right]}{r\left[\si{\micro\meter}\right]}\left[\si{\hertz}\right].
\end{equation}

A microwave resonator with resonance frequency in the $\si{\giga\hertz}$
and designed with a low impedance \citep{Lee2019} exhibits current
fluctuations comprised between $30\sim\SI{50}{\nano\ampere}$ , which
give rise to a coupling constant of only $\SI 1{\kilo\hertz}$ for
a spin situated at a distance of approximately $\SI{100}{\nano\meter}$
from the circuit.

One solution to overcome this weak coupling consists of translating
the spin degree of freedom into an electrical dipole. Indeed, the
coupling strength of an electrical dipole to a microwave resonator
may be orders of magnitude larger than for a magnetic dipole. In 2015,
Viennot et al. \citep{Viennot2015} used this method to translate
an electron spin in a carbon nanotube into an electrical dipole, by
the application of a local magnetic gradient, reaching the strong
coupling regime. Along these lines, Petta and coworkers have demonstrated
the strong coupling between a silicon double quantum dot and a microwave
coplanar waveguide resonator \citep{Mi2017}. In a follow-up experiment,
they added a cobalt micro-magnet to transfer the spin degree of freedom
to the position of the electron in the double quantum dot \citep{Mi2018}.
A spin--photon coupling rates of up to $\SI{11}{\mega\hertz}$
was reported and the strong coupling was achieved. However, due to
the presence of charge noise, the spin decoherence rate was severely
degraded to a few megahertz.

Clearly, using an electrical degree of freedom is a good solution
for reaching the strong coupling regime between a nano-object and
a microwave resonator. However, the same electrical degree of freedom
makes this nano-object sensitive to electrical noise which implies
a severe degradation of its coherence properties. A spin is intrinsically
immune to charge noise, but its magnetic coupling to a circuit is
small and thus requires large current quantum fluctuations (see Eq.
1).

In this work, we couple spins to a highly non-linear circuit with
huge current quantum fluctuations \citep{PhysRevLett.105.210501,PhysRevB.81.241202,Zhu2011}.
This circuit - called a superconducting flux qubit - consists of a
superconducting loop intersected by four Josephson junctions, among
which one is smaller than the others by a factor $\alpha$. It behaves
as a two-level system when the flux threading the loop is close to
half a flux quantum $\Phi\sim\Phi_{0}/2$ \citep{Mooij1999,PhysRevB.60.15398,Chiorescu2003}.
Each level is characterized by the direction of a macroscopic persistent
current $I_{P}$ flowing in the loop of the qubit. The value of the
persistent current $I_{P}$, typically of the order of $300-\SI{500}{\nano\ampere}$,
gives rise to a huge magnetic moment $\left(\sim\SI 5{\peta\hertz\per\tesla}\right)$,
making the energy of each level extremely sensitive to external magnetic
flux. At $\Phi=\Phi_{0}/2$, the two levels are degenerate, hybridize
and give rise to an energy splitting $\hbar\Delta$ called the flux-qubit
gap, accompanied with large current fluctuations $\delta I=I_{P}$.
Close to $\Phi_{0}/2$, the effective Hamiltonian of the circuit can
be written as

\begin{equation}
\hamil_{qb}=\frac{\hbar}{2}\left(\Delta\sigma_{qb}^{z}+\varepsilon\sigma_{qb}^{x}\right)
\end{equation}

where $\varepsilon=\frac{2I_{P}}{\hbar}\left(\Phi-\frac{\Phi_{0}}{2}\right)$,
$\sigma_{qb}^{z}$ and $\sigma_{qb}^{x}$ are Pauli operators acting
on the subspace formed by the ground $\left|0\right\rangle $ and
first excited $\left|1\right\rangle $ eigenstates of the circuit
Hamiltonian. The two major issues with flux qubit designs are device-to-device
gap reproducibility and coherence \citep{PhysRevLett.95.257002,PhysRevLett.97.167001,Bylander2011,PhysRevLett.113.123601,Yan2016}.
These problems have been recently solved by new fabrication techniques
allowing an extremely good control of e-beam lithography, oxidation
parameters of the junctions and sample surface treatment \citep{PhysRevApplied.18.064062}.

\begin{figure*}[t]
\includegraphics[width=1\textwidth]{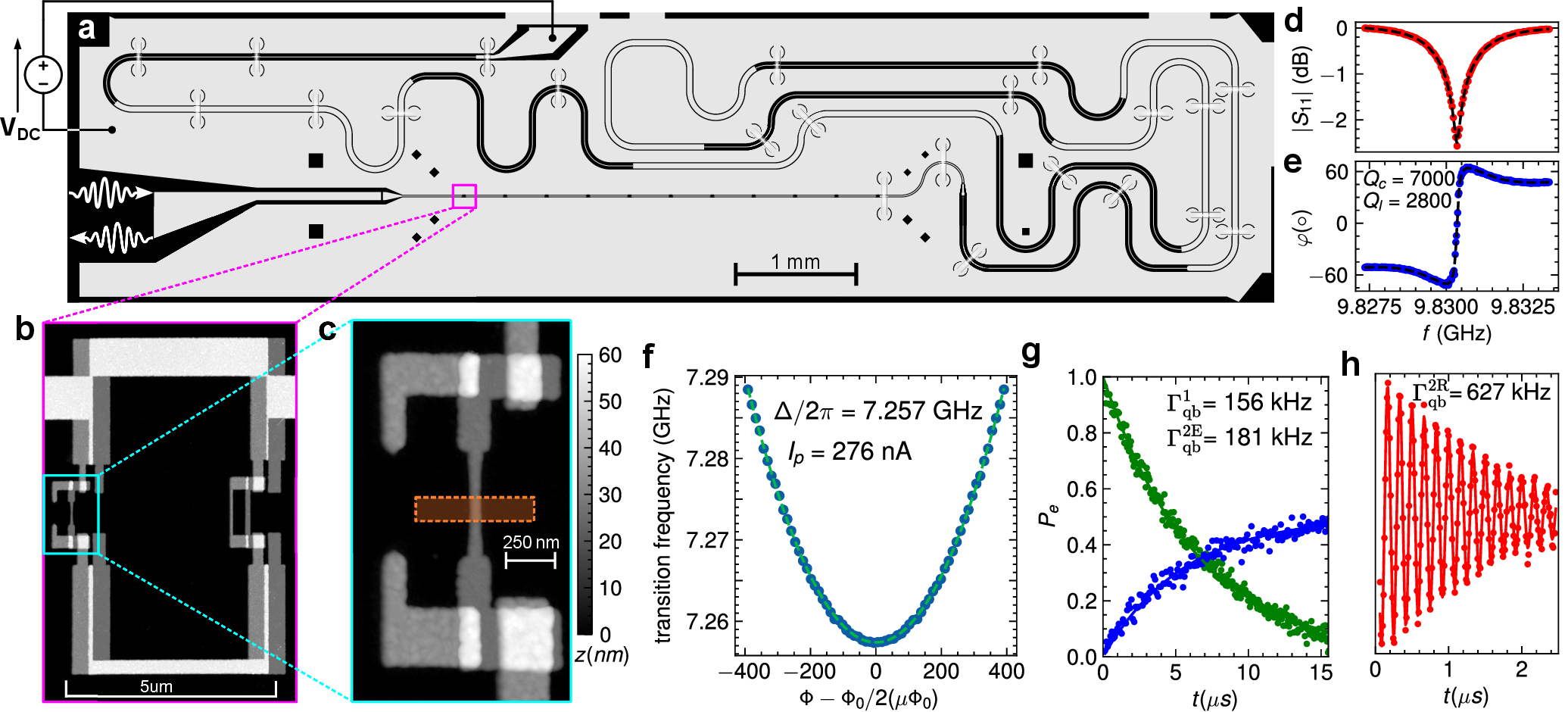}\caption{\textbf{Circuit implementation and characterization a},\textbf{ }$\lambda/2$
coplanar waveguide resonator coupled galvanically to a series of flux
qubits. The resonator is terminated on the left side by a capacitor
and on the right side by a Bragg filter (see Methods). A DC voltage is applied between
the central conductor and the ground. \textbf{b}, Atomic Force Microscope
(AFM) micrograph of one of the qubits galvanically connected to the
central conductor of the resonator. \textbf{c}, Close-up view of a
$\SI{20}{\nano\meter}$ constriction in the loop of the qubit.
This constriction is aligned with the implantation zone of $\protect\sibi$
donors represented as an orange rectangle. \textbf{d-e}, Amplitude
and phase of the reflected signal on the capacitor port.  
The total quality factor $Q_T$ of the resonator is such that $1/Q_T = 1/Q_c+1/Q_l$ , where $Q_c$ is the quality factor due to losses via the coupling capacitor 
and $Q_l$  represents the remaining losses of the circuit.
\textbf{f}, Spectroscopy of the flux qubit represented in \textbf{b} in the vicinity
of $\Phi_{0}/2$. \textbf{g}, Measurement of the relaxation of the
qubit (in green) and of the Hahn echo decay (in blue). $P_{e}$ is
the probability to find the qubit in its excited state. \textbf{h},
Measurement of the Ramsey oscillations (in red). \label{fig1}}
\end{figure*}

The second parameter to optimize the magnetic coupling is the distance
between the spin and the circuit (see Eq. 1). The spin species should
be chosen carefully to allow this precise positioning. Here, bismuth
donors in silicon (Si:Bi) are chosen since they can be implanted near
the surface with good yield and low straggling \citep{PhysRevMaterials.3.083403}
and possess long coherence times \citep{Wolfowicz2013,Ranjan2021}.
The bismuth donor has a nuclear spin $I=\frac{9}{2}$ and an electron
spin $S=\frac{1}{2}$. The Hamiltonian of a single bismuth donor in
silicon can be written as follows

\begin{equation}
\hamil_{\protect\sibi}=+\gamma_{e}\vso\cdot\vbo-\gamma_{n}\vio\cdot\vbo+\frac{A}{\hbar}\,\vso\cdot\vio
\end{equation}
The first two terms in the Hamiltonian are respectively the Zeeman
electronic and nuclear terms, $\gamma_{n}/2\pi=\SI{6.962}{\mega\hertz\per\tesla}$
being the nuclear gyro-magnetic ratio. The last term is the hyperfine
coupling term ($A/2\pi=\SI{1.4754}{\giga\hertz}$) and is isotropic
due to the symmetry of the donor. One of the advantages of bismuth
donors is their large hyperfine coupling constant, which gives rise
to an electron spin resonance transition at $5A/2\pi=\SI{7.377}{\giga\hertz}$
without application of an external magnetic field \citep{PhysRevLett.105.067601,Bienfait2015,Bienfait2016}.
As we will see in the following, this is very convenient for resonantly
coupling such a spin to a superconducting flux qubit. In this work,
the $\sibi$ donors are positioned at a depth of approximately $\SI{20}{\nano\meter}$
below the surface in regions of implantation of $500\times\SI{100}{\nano\meter}$.
Each region contains a total of approximately $60$ active electron
spins (following Poisson statistics).

Aluminum deposited directly on top of a native silicon substrate gives
rise to a Schottky barrier in which the bismuth donors may be ionized
\citep{Bienfait2015}. Indeed, aluminum has a higher work function
$\Phi=\SI{4.25}{\volt}$ than the electron affinity in silicon $\chi=\SI{4.05}{\volt}$.
Bismuth is a shallow donor situated at $E_{B,\sibi}=\SI{71}{\milli\volt}$
below the conduction band. Thus, an electron occupying the donor site
is situated at a higher electro-chemical potential than the Fermi
level of the aluminum, leading to spontaneous ionization of the donors
in the so-called depletion zone. A well-known solution to this problem
consists of introducing a thin insulating layer of silicon oxide ($\protect\siox$)
in order to prevent the exchange of electrons between the substrate
and the aluminum. Consequently, our sample is fabricated on a $\SI 5{\nano\meter}$
thermally grown silicon oxide layer \citep{PhysRevApplied.18.064062}.
By applying a positive voltage on the top electrode, it is possible
to be assured that the
donors cannot be ionized. Inversely, the donors become ionized below a specific voltage threshold. 

In Fig. \ref{fig1}a, we present a $\lambda/2$ coplanar waveguide
resonator design which allows the application of this DC voltage.
To achieve this goal, the resonator is terminated on the left by a
capacitor and on the right by a Bragg filter (see Methods). The voltage bias is
applied directly on the Bragg filter port between the central strip
and the ground plane. A positive voltage
$V=\SI{0.5}{\volt}$ should be maintained
during the cool-down, such that the donors below the central conductor
will retain their electrons (see Extended Data Fig. \ref{extfig5}). A series of eleven flux qubits is connected
galvanically to the central conductor of the resonator. In Fig. \ref{fig1}b,
we present an Atomic Force Microscope (AFM) micrograph of one of these
qubits. It contains a thin constriction of aluminum ($20\times\SI{500}{\nano\meter}$)
that crosses an implantation region as shown in Fig.\ref{fig1}c.

We now present experiments on the spectroscopic measurements for characterization
of the qubit-resonator system (See Methods for more details on the
experimental setup). In Fig \ref{fig1}d-e, we present the amplitude
and phase of the reflected signal on the capacitor port. One can extract
from these measurements the frequency of the resonator $\omega_{r}/2\pi=\SI{9.832}{\giga\hertz}$
and its quality factor $Q=2000$. In Fig. \ref{fig1}f, we present
the spectroscopy of the flux qubit which will be used in the following
experiment to detect and couple $\sibi$ spins. One can extract from
this measurement the qubit gap $\Delta/2\pi=\SI{7.257}{\giga\hertz}$
and its persistent current $I_{p}=\SI{276}{\nano\ampere}$. We now
turn to the coherence times at the so-called optimal point where the
qubit frequency is minimal and thus insensitive to flux-noise to first
order \citep{PhysRevLett.95.257002,PhysRevLett.97.167001,PhysRevApplied.18.064062}.
Energy relaxation decay is shown in Fig. \ref{fig1}g from which we
extract $\Gamma^{1}_{qb}\sim\SI{150}{\kilo\hertz}$. This decay rate is
considerably higher than what was obtained in Ref. \citep{PhysRevApplied.18.064062}
and may be due to higher dielectric losses in the substrate due to
the presence of residual dopants or to a bad interface between the epilayer and the substrate. Ramsey fringes show an {exponential}
decay with $\Gamma^{2R}_{qb}\sim\SI{630}{\kilo\hertz}$ and pure dephasing
rate $\Gamma^{\varphi R}_{qb}=\Gamma^{2R}_{qb}-\Gamma^{1}_{qb}/2\sim\SI{550}{\kilo\hertz}$.
The spin-echo decays {exponentially} with a pure echo dephasing rate of $\Gamma^{\varphi E}_{qb}\sim\SI{100}{\kilo\hertz}$.
Away from the optimal point the pure dephasing rate becomes dominated
by 1/f flux noise and increases proportionally with $\varepsilon$,
giving access to the amplitude of flux noise ${\sqrt{A}=2.0\:\mathrm{\mu\Phi_{0}}}$
\citep{PhysRevApplied.18.064062}. This value is significantly higher
than typical flux noise amplitudes for qubits fabricated with the
same technique and measured under similar conditions \citep{PhysRevApplied.18.064062}
and again may be related to the presence of residual dopants in the
substrate.

\begin{figure*}[t]
\includegraphics[width=1\textwidth]{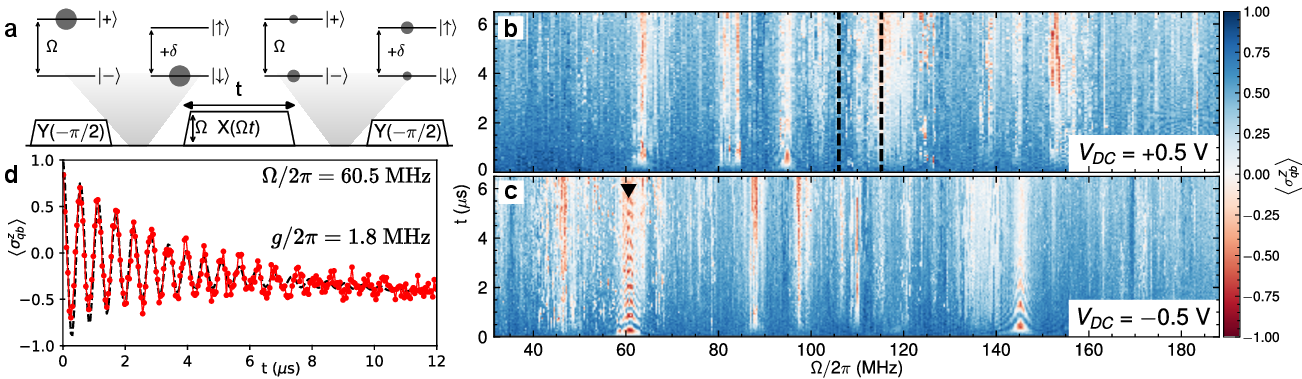}\caption{\textbf{Detection of two-level systems by spin-locking a}, A first
pulse $Y(-\pi/2)$ puts the flux qubit in its $\left|+\right\rangle $
state. A second pulse $X(\Omega t)$ tunes on the interaction of the
qubit with a two-level system when the condition $\Omega=\left |\delta \right |$
is satisfied. A third pulse $Y(-\pi/2)$ projects the flux qubit
to its excited state $\left|1\right\rangle $ in the absence of resonant
interaction. \textbf{b}, Expectation value $<\sigma_{qb}^{z}>$
of the flux qubit state, measured after the pulse sequence shown in
\textbf{a}, versus pulse amplitude $\Omega$ and duration $t$ when
$V_{DC}=\SI{0.5}{\volt}$. A close-up view of the region surrounded by the black dashed lines in  shown in Fig. \ref{fig3}b.  \textbf{c}, Expectation value
$<\sigma_{qb}^{z}> $ of the flux qubit state
as a function of pulse amplitude $\Omega$ and duration $t$, when
$V_{DC}=\SI{-0.5}{\volt}$. \textbf{ d}, Signal measured
at $V_{DC}=\SI{-0.5}{\volt}$ and $\Omega/2\pi = \SI{60.5}{\mega\hertz}$, indicated by a black arrow in\textbf{ c}.
A two-level system (TLS) of frequency $\omega_{s}/2\pi=\left(\Delta+\Omega\right)/2\pi=\SI{7.3175}{\giga\hertz}$
is detected. The dashed line is the result of a Linblad simulation assuming jump operators 
$L^1_{qb,s}=\sqrt{\Gamma^1_{qb,s}}\sigma^-_{qb,s}$  and
$L^\varphi_{qb,s}=\sqrt{\Gamma^\varphi_{qb,s}}\sigma^z_{qb,s}$ for the flux qubit and the TLS respectively. The value of $\Gamma^1_{qb} = \SI{150}{\kilo\hertz}$ and 
$\Gamma^\varphi_{qb} = \SI{40}{\kilo\hertz}$ are taken from the qubit characterization. A fit of the experimental data gives us $\Gamma^1_{s} = \SI{140}{\kilo\hertz}$, $\Gamma^\varphi_{s} = \SI{200}{\kilo\hertz}$ and the coupling between the flux qubit and the TLS $g/2\pi=\SI{1.8}{\mega\hertz}$.  \label{fig2}}
\end{figure*}

Clearly, working at the flux qubit optimal point is required if one
wishes to have a flux qubit with good coherence properties. It is
therefore important to engineer the flux qubit gap to be resonant
with the spins if one wishes to make a coherent exchange. In this
work, we achieve this condition by applying a resonant Rabi drive
on the flux qubit biased at its optimal point \citep{PhysRevA.92.052335}.
The driven Hamiltonian is written as

\begin{equation}
\hamil=\hbar\frac{\Delta}{2}\sigma_{qb}^{z}+\frac{\hbar\omega_{s}}{2}\sigma_{s}^{z}+\hbar g\sigma_{qb}^{x}\sigma_{s}^{x}+\hbar\Omega\sigma_{qb}^{x}\cos\left(\Delta t\right)
\end{equation}
Even if the coupling constant $g$ is several orders of magnitude
smaller than the detuning $\delta=\omega_{s}-\Delta$ , one can show
that this Hamiltonian is equivalent to a time-independent flip-flop
Hamiltonian when the condition $\Omega=\left |\delta \right |$ is satisfied (for
more details see Methods). The advantage of this coupling scheme is
that one can turn on and off the coupling by controlling the amplitude
$\Omega$ of the microwave tone.

In Fig. \ref{fig2}a, we present a protocol to detect the presence
of impurity donors or more generally any two-level systems (TLS) in
the vicinity of the flux qubit. A first pulse operates a rotation
of $-\pi/2$ around the Y axis of the Bloch sphere and transfers the
qubit initialized in its ground state to the $\left|+\right\rangle =\left(\left|0\right\rangle +\left|1\right\rangle \right)/\sqrt{2}$
state. This state being an eigenstate of $\hbar\Omega\sigma_{qb}^{x}/2$
should remain unchanged in the presence of a second pulse along the X axis of amplitude
$\Omega$ and frequency $\Delta$. The third pulse operates a rotation
of $-\pi/2$ around the Y axis and brings the qubit to its excited
state, $\left|1\right\rangle $. The presence of an impurity donor
or TLS modifies this picture when $\Omega=|\delta|$ \citep{PhysRevB.102.100502}. In that case,
the qubit can exchange its excitation during the application of the
second pulse with a spin or TLS initially in its ground state $\left|\downarrow\right\rangle $.
Consequently, the flux qubit state does not reach $\left|1\right\rangle $
after the third pulse is applied.

Fig. \ref{fig2}b, shows a color plot representing the flux qubit
state at the end of the sequence versus the second pulse amplitude
and duration. A change in the color is observed when the qubit can
efficiently exchange energy with a surrounding TLS. The measured spectrum
unveils that a large number of TLSs interacts with the qubit. This
spectrum may vary as a function of time on a typical timescale
hours or days. Some rare events affect several TLSs together with
frequency jumps that can reach tens of megahertz. The spectrum can
also be modified by changing the gate voltage as shown in Fig. \ref{fig2}c.
If the TLS is an impurity donor in the silicon wafer, it should be
ionized when a negative voltage is applied. Some TLSs are strongly
coupled to the qubit and can exhibit coherent exchange (see Fig. \ref{fig2}d)
but most of them have a rather short relaxation time.

In the following, we will use this short relaxation time as a way
to filter the signal coming from the $\sibi$ donors. Indeed, the
relaxation time $T_{1}$ of donors in silicon can be extremely long
at low temperatures. In Ref. \citep{Pla2012}, the relaxation time
of a single phosphorous donor was measured to be $T_{1}\sim\SI{0.7}{\second}$.
In Ref. \citep{Bienfait2016} non-radiative energy relaxation of an
ensemble of bismuth donors was measured to be even longer, reaching
approximately $\SI{1500}{\second}$ at dilution temperatures $T=\SI{20}{\milli\kelvin}$.

\begin{figure*}[t]
\includegraphics[width=1\textwidth]{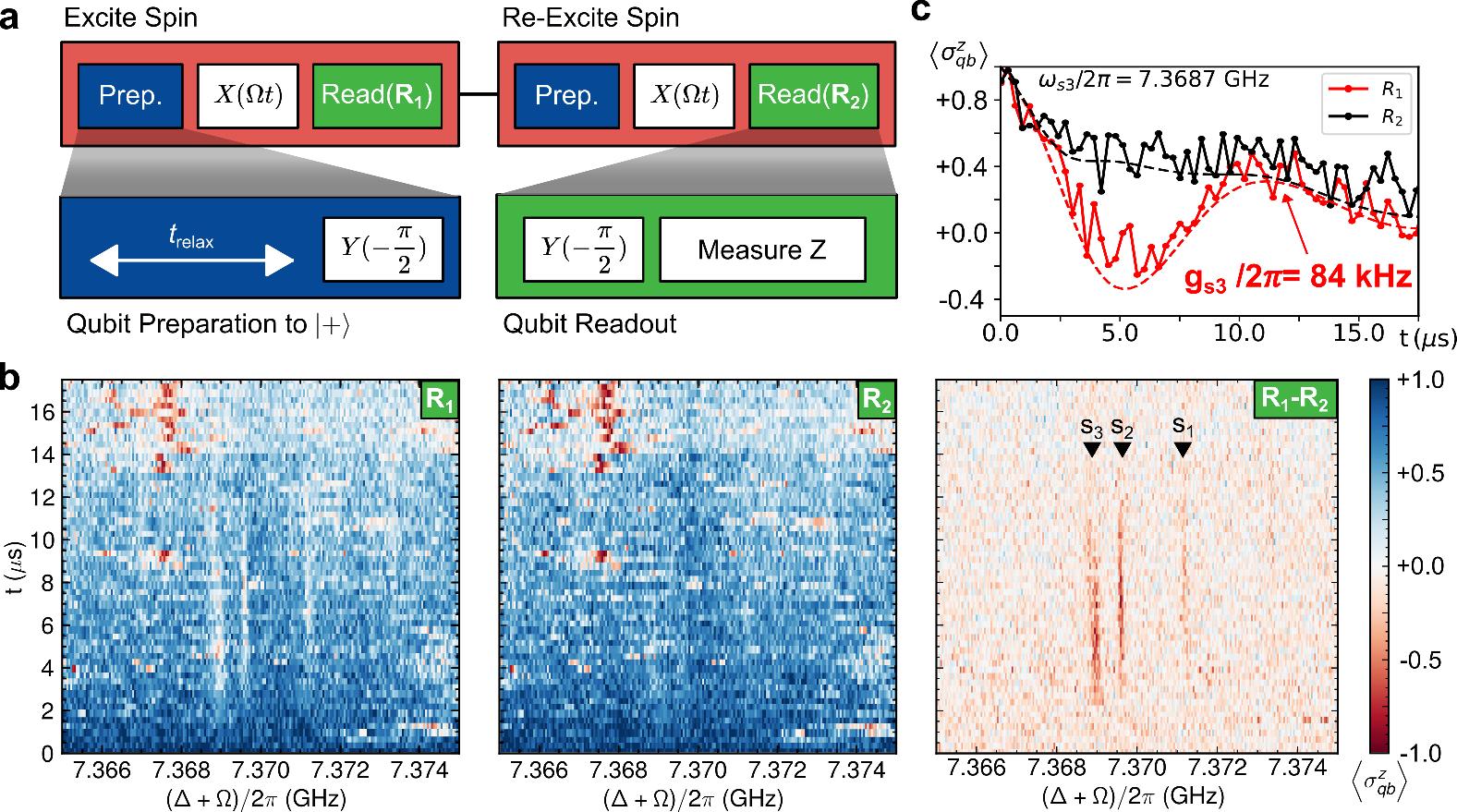}\caption{\textbf{Detection of Bismuth donors.} 
\textbf{a}, Protocol for filtering out two-level systems with short relaxation times ($T_{s}^1\ll t_{\text{relax}}$). To achieve this, we repeat the sequence shown in  Fig.\ref{fig2}.a  and compare the results of the first ($R_1$) and second ($R_2$) qubit readouts.  \textbf{b},  Qubit state readout versus Rabi frequency ($\Omega$) and interaction time ($t$) after the first pulse sequence ($R_1$) , after the second pulse sequence ($R_2$) and the difference between the readout results ($R_1-R_2$) . Only spins with a relaxation time longer than the relaxation time $t_{\text{relax}}$ are still visible.
Three $\sibi$ donors are detected at $\omega_{s1}/2\pi=\SI{7.3712}{\giga\hertz}$,
$\omega_{s2}/2\pi=\SI{7.3692}{\giga\hertz}$, $\omega_{s3}/2\pi=\SI{7.3687}{\giga\hertz}$.
These three spectral lines disappear when the bias voltage is tuned
to $V_{DC}=\SI{-0.5}{\volt}$. 
\textbf{c}, Coherent oscillations
between the driven flux qubit and spin 3.  The dashed lines are the results of a Linblad simulation assuming  jump operators 
$L^1_{qb}=\sqrt{\Gamma^1_{qb}}\sigma^-_{qb}$  and
$L^\varphi_{qb}=\sqrt{\Gamma^\varphi_{qb}}\sigma^z_{qb}$ 
for the  flux qubit, with 
$\Gamma^1_{qb} = \SI{150}{\kilo\hertz}$ and 
$\Gamma^\varphi_{qb} = \SI{40}{\kilo\hertz}$ and assuming no decoherence or relaxation from the bismuth spin.
From this measurement, one
can extract the coupling constant between the qubit and the spin $g_{s3}/2\pi=\SI{84}{\kilo\hertz}$.
\label{fig3}}
\end{figure*}

In Fig. \ref{fig3}a-b, we apply twice the protocol described in Fig.
\ref{fig2}a and represent the difference between the measurement
after the first and second pulse sequences. Thus, only species with
a relaxation time longer than the duration of the sequence will appear.
In our range of interest, three lines can still be observed. We measure
signals at $\omega_{s1}/2\pi=\SI{7.3712}{\giga\hertz}$, $\omega_{s2}/2\pi=\SI{7.3692}{\giga\hertz}$,
$\omega_{s3}/2\pi=\SI{7.3687}{\giga\hertz}$, which disappear under
a negative voltage bias. 

As mentioned earlier, most random TLSs have a rather short relaxation time in the few microsecond range \citep{PhysRevLett.105.177001,Lisenfeld2016}. However, some rare TLSs may exhibit longer relaxation times \citep{Spiecker2023}. To address this concern, we performed a statistical analysis of detectable TLSs (see Extended Data Fig. \ref{extfig4}). For a time $t_\text{relax}=\SI{16}{\micro\second}$, the spectral density of these long-lived TLSs is around $\SI{0.22}{\mega\hertz}^{-1}$. As we increase $t_\text{relax}$ ($\SI{32}{\micro\second}$ or $\SI{60}{\micro\second}$), the spectral density of TLSs  decreases to $\SI{0.17}{\mega\hertz}^{-1}$ or $\SI{0.09}{\mega\hertz}^{-1}$ respectively. At $t_\text{relax}=\SI{150}{\micro\second}$, the TLSs become very rare with a spectral density of $\SI{0.06}{\mega\hertz}^{-1}$.

With approximately 60 activated spins intentionally introduced in the implantation region, we expect to find around 3 spins beneath the aluminum constriction. The probability to detect three random TLSs with long relaxation time in a range of 4 MHz exists but is very low (less than $2\%$). Additionally, the behavior of the detected spectral line under the application of an electric field (see Extended Data Fig. \ref{extfig5}) is in excellent agreement with the expected Stark shift of bismuth donors \citep{PhysRevB.90.195204}.

The frequencies of the detected signals are close to the zero field
splitting of bismuth donors in bulk silicon (7.377 GHz) but slightly
shifted. The shift of the spectral lines is due to the close proximity of the aluminum
wire, which generates mechanical stress resulting from the mismatch
of the coefficient of thermal expansion between the substrate and
the metal. This stress induces a shift in the hyperfine coupling strength
that has been experimentally measured \citep{PhysRevLett.120.167701,PhysRevApplied.9.044014}.
To first approximation, one may introduce modifications to the constant
$A$ depending on diagonal terms only of the strain tensor $\varepsilon$.
Namely,
\begin{equation}
\frac{\Delta A}{A}=\frac{K}{3}\left(\varepsilon_{xx}+\varepsilon_{yy}+\varepsilon_{zz}\right)
\end{equation}
 with $K=19.1$ \citep{PhysRevApplied.9.044014}. Using the simulation
of Extended Data Fig. \ref{extfig2}c, we see that these frequency shifts are compatible
with bismuth donors lying in the close vicinity of the constriction
and we can infer their approximate distance to the circuit.

In Fig. \ref{fig3}c, we present the expectation value $<\sigma_{qb}^{z}> $
as a function of interaction time $t$ between one of the donors (spin
s3) and the resonantly driven qubit. Coherent exchange between the
driven flux qubit and the bismuth electron spin is observed. This
measurement enables us to extract the coupling between the qubit and
the spin, as was done previously in Fig. \ref{fig2}c for an arbitrary
TLS. We repeat this measurement for the three single donors and find
their respective coupling constants with the qubit, namely $g_{s1}/2\pi\sim\SI{45}{\kilo\hertz}$,
$g_{s2}/2\pi\sim\SI{62}{\kilo\hertz}$ and $g_{s3}/2\pi\sim\SI{84}{\kilo\hertz}$.
These values are in good agreement with what is expected from a simple Biot
and Savart simulation (see Extended Data Fig. \ref{extfig2}d.), assuming that spins are point-like particles. 

A first application of the coupling described herein above consists
of quickly initializing the spins, without waiting for their long
relaxation time. Preparing the state of the flux qubit in $\left|+\right\rangle $
(or $\left|-\right\rangle $) and letting the system evolve for $t=\pi/g$
will set the spin to state $\left|\uparrow\right\rangle $ (or $\left|\downarrow\right\rangle $
respectively), independently of the initial spin state. In our experiments,
the purity of the spin state is affected by the flux qubit decoherence
during the interaction time. One way to improve the preparation of
the target state ($\left|\downarrow\right\rangle $ or$\left|\uparrow\right\rangle $)
consists of repeating the protocol typically two to five times \citep{PhysRevA.92.052335}.
The repetition of the protocol improves the state initialization until
it reaches an asymptotic value.

\begin{figure*}[t]
\includegraphics[width=1\textwidth]{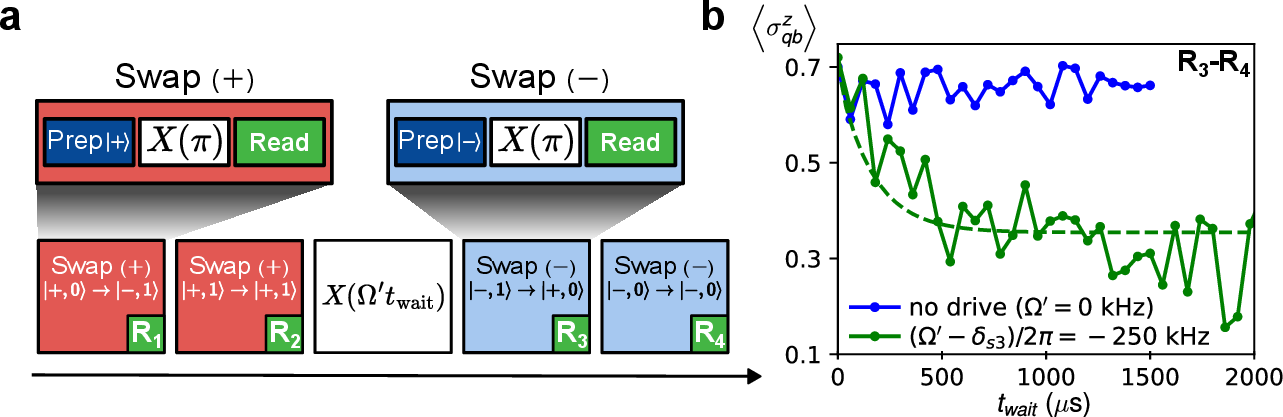}\caption{\textbf{Initialization and Purcell relaxation of the spin.} \textbf{a}, Protocol for initializing the spin state in its excited state state (Swap$(+)$) or in its ground state (Swap$(-)$) and for measuring the spin relaxation time. The purity of the spin state after a simple Swap is around $75\%$  but it may be increased to around $82\%$ by repeating 3-5 times the protocol. Higher purity could be obtained if the flux qubit had longer relaxation time. \textbf{b}, Measurement of the qubit state $<\sigma^z_{qb}>$ versus waiting time $t_{\text{wait}}$ following the protocol shown in  \textbf{a}. Without Rabi drive ($\Omega'=0$), the intrinsic relaxation rate of the spin is much smaller than the measurement timescales ($\Gamma^1_{s}\leq\SI{20}{\hertz}$). In presence of a Rabi drive ($(\Omega'-\delta_{s3})/2\pi=-\SI{250}{\kilo\hertz}$), the spin relaxation rate is increased to $\Gamma^P_{s}\sim\SI{0.3}{\kilo\hertz}$. The dashed line corresponds to the results of a Lindblad simulation using the parameters introduced in Fig. \ref{fig3}c. 
\label{fig4}}
\end{figure*}

Once its state is well initialized, the spin can serve as a quantum
memory for the flux qubit. In Fig. \ref{fig4}a-b, we present the simplest
case illustrating a possible use of this memory. Here, the spin is
first initialized in $\left|\uparrow\right\rangle $. After a waiting
time $t_{wait}$, its state is mapped back to the qubit by letting
the system evolve for $t=\pi/g$ with a qubit prepared initially
in $\left|-\right\rangle $ state. This measurement should enable
us to extract information about the relaxation time of the spin. This
time is very long, much larger than the millisecond scale on which
our measurement is done. In order to increase the relaxation rate
of the spin to some reasonable values, one can drive the qubit out
of resonance during the waiting time in order to open a relaxation
channel by Purcell effect via the qubit \citep{Bienfait2016}. For
instance, the relaxation rate of the spin is increased to $\Gamma^P_{s}\sim\SI{0.3}{\kilo\hertz}$
when the qubit is detuned from the spin by $(\Omega-\delta_{s3})/2\pi=\SI{-250}{\kilo\hertz}$.

The fidelity of the quantum memory protocol can be increased by having
a qubit with longer relaxation and dephasing times. We suspect that
the relatively short relaxation time of our qubit is related to the presence of residual dopants or a to
bad interface between the isotopically purified epilayer and the substrate.
In future experiments, we will rather use a different technique where
$\sipure$ isotopically enriched regions can be directly formed on
an intrinsic high resistivity sample. This technique allows for local
isotopic enrichment through sputtering and implanting of a high fluence
of $\sipure$ ions \citep{PhysRevMaterials.5.014601}. Another important
parameter to improve in further experiments is the stability of the
drive power, which currently represents an intrinsic limitation of
our experimental technique. Indeed, a typical microwave power stability
of 0.2\% will give rise to variations of the Rabi frequency of $\delta\Omega\sim\SI{100}{\kilo\hertz}$,
which is an issue for obtaining a good coherent exchange. One possible
way to solve this question consists of using a double drive scheme
where the flux qubit is driven simultaneously along two axes. This
kind of technique has been already demonstrated for stabilizing Rabi
oscillations \citep{PhysRevA.96.013850}. It could be possible to
exploit it further in order to establish an effective coupling between
two spins, using the qubit as a mediator.


\section{Methods} 

\subsection{Experimental Setup}

Experiments are performed at a temperature of $\sim\SI{15}{\milli\kelvin}$
in a dilution refrigerator. Extended Data Fig. \ref{extfig1} shows
a detailed schematic of the experimental setup. The sample is glued
on a microwave printed circuit board made out of TMM10 ceramics, then
enclosed in a copper box with low mode volume which is itself embedded
into a superconducting coil that is used to provide magnetic flux
biases to the qubits. To reduce low frequency magnetic noise, the
coil is surrounded by a superconducting enclosure and magnetically
shielded with a high permeability metal box. The apertures of the
box are tightly closed using microwave absorber, in order to protect
the sample from electromagnetic radiation that could generate quasiparticles.

The coil is powered by a voltage source filtered by a custom designed
ultra-stable voltage to current converter. The microwaves are generated
by ultra-low noise analog microwave synthesizers. The pulses are modulated
at an intermediate frequency of {$10-\SI{100}{\mega\hertz}$}
by a Quantum Machines OPX system connected to IQ mixers. Voltage controlled
attenuator is used to adjust the measurement pulse amplitude. The
input line is attenuated at 4K stage (-20 dB) and {at
the mixing chamber stage (-30 dB)}to minimize
thermal noise and filtered with homemade impedance-matched copper
powder filters (-3 dB @ {$\SI{10}{\giga\hertz}$}).
In addition, the pulses are shaped with smooth rise and fall ($\sim20$
ns) in order to reduce the population of microwave photons in the
resonator during coherent state evolution of the qubit. A DC voltage ($V_{DC}$)
is applied on the Bragg filter port of the resonator using a bias
tee thermalized at $\SI{15}{\milli\kelvin}$. Additionally, the input
noise on this port is strongly filtered by an homemade Eccosorb low-pass
filter (-150 dB @ {$\SI{10}{\giga\hertz}$).}

Qubit state measurement is done using dispersive readout, by measuring
the reflection of microwave pulses on the capacitor port of the resonator,
using a custom built setup. {The readout output line
is filtered by two shielded double circulators and a $8-\SI{12}{\giga\hertz}$
band pass filter.} The readout output signal is amplified using a
low-noise cryogenic HEMT amplifier and a room temperature amplifier.
After demodulation, the quadratures of the readout output pulse are
sampled and averaged using the IQ inputs of the OPX system. At this
point, we perform a principal axis transformation on the data points
by diagonalizing their covariance matrix. Using this transformation,
we extract the largest principal component of the measured $(I,Q)$
points and obtain the state of the qubit.

\subsection{Sample design and fabrication}

The sample is fabricated on a $\SI{300}{\micro\meter}$ thick wafer
of natural silicon covered by a $\SI{10}{\micro\meter}$ isotopically
enriched epilayer containing 730 ppm of residual $\sibad$ and background
doping of approximately $3\times10^{16}\:\si{\per\centi\meter\cubed}$
phosphorus donors. Before implantation, the sample is covered by a
thermally grown $\SI 5{\nano\meter}$ width $\protect\siox$ layer. This
gate oxide step is followed by a forming gas annealing to passivate
dangling bounds at the {$\protect\siox/\silicon$ interface.}

\subsubsection{Bismuth donors implantation}

The $\sibi$ donors are located in regions of size $500\times\SI{100}{\nano\meter}$
which are defined by e-beam lithography using a polymethyl methacrylate
(PMMA) mask ($\SI{200}{\nano\meter}$). The $\sibi$ donors are positioned
at a depth of approximately $\SI{20}{\nano\meter}$ below the surface
by ion implantation with an energy of 26 keV (see SRIM simulation
in Extended Data Fig. \ref{extfig2}). The implantation area density
is $2\cdot10^{11}\si{\per\centi\meter\squared}$, which translates
into a peak density of bismuth impurities of $2\cdot10^{17}\si{\per\centi\meter\cubed}$.
The implantation is followed by a cleaning step to remove the PMMA
mask and a rapid thermal annealing ($\SI{1000}{\celsius}$, Ar, $\SI 5{\second}$)
to activate the bismuth donors. Taking into account an activation
ratio of 60\% \citep{PhysRevMaterials.3.083403}, a total of approximately
60 electron spins are present in each implantation region. {Dangling
bounds at the $\protect\siox/\silicon$ interface are passivated with an evaporation
of a}$\SI{50}{\nano\meter}$ aluminum layer followed
by a forming gas annealing ($\SI{400}{\celsius},\SI{30}{\minute})$.
This aluminum layer acts as a catalyst for atomic hydrogen which efficiently
passivates the interface states. Then, the sample is etched by aluminum
etchant and cleaned.

\subsubsection{Coplanar waveguide resonator fabrication}

After one night of pumping, we evaporate a clean layer of $\SI{150}{\nano\meter}$
aluminum onto the chip. Optical resist (AZ1505) is spun on the sample
and a $\lambda/2$ coplanar waveguide resonator is patterned by UV
laser lithography. This resonator is terminated on the left side by
a capacitor and on the right side by a Bragg filter. The value of
the capacitance is calculated by an electromagnetic simulator (Sonnet)
to be $C_{C}\sim\SI 5{\femto\farad}$. The Bragg filter is formed
by eleven layers of identical length $L_{BF}=\SI{3.44}{\milli\meter}$
with alternate characteristic impedance $Z_{1}=\SI{35}{\ohm}$ and
$Z_{2}=\SI{80}{\ohm}$. This change of characteristic impedance is
obtained by changing the width of the central conductor and its spacing
to the ground. The Bragg filter is designed to have a band gap centered
around $\SI{8.3}{\giga\hertz}$  with a bandwidth at -3 dB of +/-  $\SI{2.4}{\giga\hertz}$ , in order to protect the qubits from relaxation via the line. After development, the wafer is etched with
aluminum etchant, followed by cleaning in N-Methylpyrrolidone (NMP)
overnight.

\subsubsection{Flux qubit fabrication}

The flux qubits are fabricated by e-beam lithography using the so-called
Dolan technique. In this work, we used a trilayer process \citep{PhysRevLett.113.123601,PhysRevApplied.18.064062},
in which a germanium mask, suspended on top of a $\sim\SI{650}{\nano\meter}$
thick copolymer resist is employed. To achieve this, we spin a bilayer
of copolymer resist (MMA(8.5)MAA-EL7), evaporate $\SI{60}{\nano\meter}$
of germanium onto the chip and spin a high contrast e-beam resist
(CSAR 62) on the top of the germanium layer. The qubits are patterned
by electron-beam lithography ($\SI{50}{\kilo\volt}$, $\SI{660}{\micro\coulomb\per\centi\meter\squared}$).
The development takes place in a solution of methyl isobutyl ketone/
isopropanol (MIBK/IPA=3:1) for 240s, followed by 60s in IPA. The chip
is then loaded into a Reactive Ion Etcher (RIE) to perform plasma
etching with sulphur hexafluoride ($\mathrm{SF_{6}}$) in order to
form the rigid germanium mask. The rigidity and conductance of the
germanium mask helps to achieve sharper resolution and more stable
dimensions. Moreover, this mask is immune to oxygen ashing which allows
for cleaning the region below the mask.

After the RIE process, the copolymer resist that has been exposed
under e-beam lithography is developed using MIBK/IPA=3:1. Areas under
the qubit are thus automatically cleaned all the way down to the substrate
surface. A final step of oxygen ashing further cleans the regions
where aluminum will be deposited.

The sample is then loaded into a Plassys MEB 550S electron-beam evaporator
and pumped overnight. For achieving the best resolution, we evaporate
a first aluminum layer of $\SI{25}{\nm}$ perpendicularly onto the
sample ($\theta_{1}=\SI 0{\degree}$). This is done at low temperature
($\sim\SI{-50}{\celsius}$) to enable fine aluminum grain size required
for the realization of the constrictions. A dynamical oxidation of
the first aluminum layer is performed, by introducing a dynamic flow
of oxygen/argon (15\%-85\%) at a pressure of $\SI{20}{\micro\bar}$
for 30 minutes. A second layer of $\SI{30}{\nano\meter}$ of aluminum
is then evaporated at a temperature of $\SI{\sim7}{\celsius}$ with
an angle $\theta_{2}\sim\SI{35}{\degree}$. This angle is adjusted
in order to obtain a displacement $d=\SI{420}{\nano\meter}$ with
respect to the first image. Finally, we make a static oxidation at
pressure of $10\,\mathrm{mbar}$ for 10 minutes. This last step encapsulates
the junctions with aluminum oxide and allows for a better aging.

\subsubsection{Ion milling recontact}

Direct contact between the coplanar waveguide resonator and the flux
qubit does not form conductive contact due to the native oxide formed
between the fabrication steps. To establish a galvanic contact, we
re-connect the qubit to the resonator by evaporating a $\sim\SI{150}{\nano\meter}$-thick
aluminum on the overlap regions, with an ion milling step (Argon,
$V_{emitter}=\SI{500}{\volt}$, $I_{anode}=\SI{17.5}{\milli\ampere}$,
$\SI{20}{\second}$) prior to evaporation in order to remove the
native oxide.

\subsubsection{Characterization of the constrictions}

The width of the evaporated aluminum constrictions should be minimal
to increase the coupling. As shown in Extended Data Fig. \ref{extfig3}, constrictions
may manifest a penumbra effect when the germanium mask slit is sufficiently
thin. This occurs when the viewing angle of the extended source is
larger than the angle of the slit opening seen from the substrate
surface. By conservation of matter, the quantity of evaporated aluminum
passing through the slit is always given by $s\times t$ where $s$
is the width of the slit and $t$ is the thickness of evaporated aluminum.
However, the aluminum is evaporated over a broadened zone on the substrate,
resulting in partially evaporated regions (\emph{penumbra} \textit{effect}).
The resulting profile can be computed as a convolution of the slit
{with the gaussian distribution of the source width.
}In Extended Data Fig. \ref{extfig3}b, we show the broadened profile evaporated
on the substrate as a function of the slit width, assuming $t=\SI{25}{\nano\meter}$,
coming from a source with a gaussian distribution of diameter $W=\SI{10}{\milli\meter}$
at a distance $D=\SI{500}{\milli\meter}$ and a mask suspended at
height $h=\SI{650}{\nano\meter}$. The maximum thickness of the profile
grows with the slit width (see Extended Data Fig. \ref{extfig3}c) and saturates
only above a certain slit size $s\approx\SI{25}{\nm}$. The minimal
width is $\approx\SI{13}{\nm}$. We show an AFM micrograph of a typical
constriction. The height profile of the constriction along the cut
indicated in Extended Data Fig. \ref{extfig3}d is shown in Extended Data Fig. \ref{extfig3}e
and illustrates the penumbra effect.

\subsection{Dynamical coupling between the flux qubit and an arbitrary two-level
system}

In this work, we use a coupling scheme which can be turned on and
off by the application of a microwave tone. In the spirit of Ref.
\citep{PhysRevA.92.052335}, we consider that the flux qubit is biased
at its optimal point and coupled to a two-level system by a coupling
constant $g$ which is several orders of magnitude smaller than the
detuning $\delta=\omega_{s}-\Delta$ between the resonance frequencies
of the two systems. One applies a resonant Rabi drive on the flux
qubit. The driven Hamiltonian can be written as

\[
\hamil=\hbar\frac{\Delta}{2}\sigma_{qb}^{z}+\frac{\hbar\omega_{s}}{2}\sigma_{s}^{z}+\hbar g\sigma_{qb}^{x}\sigma_{s}^{x}+\hbar\Omega\sigma_{qb}^{x}\cos\left(\Delta t\right)
\]
Under unitary transformation $U_{1}=\exp\left(\im\frac{\Delta}{2}(\sigma_{qb}^{z}+\sigma_{s}^{z})t\right)$
and after rotating wave approximation, we get
\[
\hamil_{1}=\hbar\frac{\delta}{2}\sigma_{s}^{z}+\hbar\frac{\Omega}{2}\sigma_{qb}^{x}+\hbar g\left(\sigma_{qb}^{+}\sigma_{s}^{-}+\sigma_{qb}^{-}\sigma_{s}^{+}\right)
\]
 In the eigenbasis $\ket{\mp}$ of $\hbar\Omega/2\sigma_{qb}^{x}$,
the above operators can be replaced by
\begin{align*}
\sigma_{qb}^{\pm} & \to\left(\sigma_{qb}^{z}\mp\im\sigma_{qb}^{y}\right)/2\\
\sigma_{qb}^{x} & \to\sigma_{qb}^{z}
\end{align*}
In this basis $\hamil_{1}$ can be written as
\begin{align*}
\hamil_{1} & =H_{0}+V\\
H_{0} & =\hbar\Omega\frac{\sigma_{qb}^{z}}{2}+\hbar\delta\frac{\sigma_{s}^{z}}{2}\\
V & =\hbar g\left(\frac{\sigma_{qb}^{z}-\im\sigma_{qb}^{y}}{2}\sigma_{s}^{-}+\frac{\sigma_{qb}^{z}+\im\sigma_{qb}^{y}}{2}\sigma_{s}^{+}\right)
\end{align*}

The expression of operators $\sigma_{1}^{-}$, $\sigma_{1}^{+}$,
$\sigma_{2}^{-}$, $\sigma_{1}^{+}$ under unitary transformation
$U_{2}=\exp\left(\im(\frac{\Omega}{2}\sigma_{qb}^{z}+\frac{\delta}{2}\sigma_{s}^{z})t\right)$
can be easily estimated using Baker Campbell Hausdorff formula

\begin{eqnarray*}
\sigma_{qb}^{+} & \to & \sigma_{qb}^{+}e^{+\im\Omega t}\\
\sigma_{qb}^{-} & \to & \sigma_{qb}^{-}e^{-\im\Omega t}\\
\sigma_{s}^{+} & \to & \sigma_{s}^{+}e^{+\im\delta t}\\
\sigma_{s}^{-} & \to & \sigma_{s}^{-}e^{-\im\delta t}
\end{eqnarray*}
Therefore under this transformation, the Hamiltonian becomes

\begin{equation*}
\hamil_{2}=U_{2}VU_{2}^{+}=\hbar g\left(\frac{\sigma_{qb}^{z}+\sigma_{qb}^{-}e^{-\im\Omega t}-\sigma_{qb}^{+}e^{+\im\Omega t}}{2}\sigma_{s}^{-}e^{-\im\delta t}+\frac{\sigma_{qb}^{z}+\sigma_{qb}^{+}e^{+\im\Omega t}-\sigma_{qb}^{-}e^{-\im\Omega t}}{2}\sigma_{s}^{+}e^{+\im\delta t}\right)
\end{equation*}

If $\Omega=\delta$, only two terms of this Hamiltonian will be time
independent, giving rise to an effective Hamiltonian

\begin{align*}
H_{\Omega=\delta} & =-\frac{\hbar g}{2}\left(\sigma_{qb}^{+}\sigma_{s}^{-}+\sigma_{qb}^{-}\sigma_{s}^{+}\right)
\end{align*}


\renewcommand{\figurename}{Extended Data Fig.}
\renewcommand{\thefigure}{\arabic{figure}}
\renewcommand{\theHfigure}{Ex\arabic{figure}}
\setcounter{figure}{0}

\begin{figure*}[p]
\includegraphics[width=\textwidth]{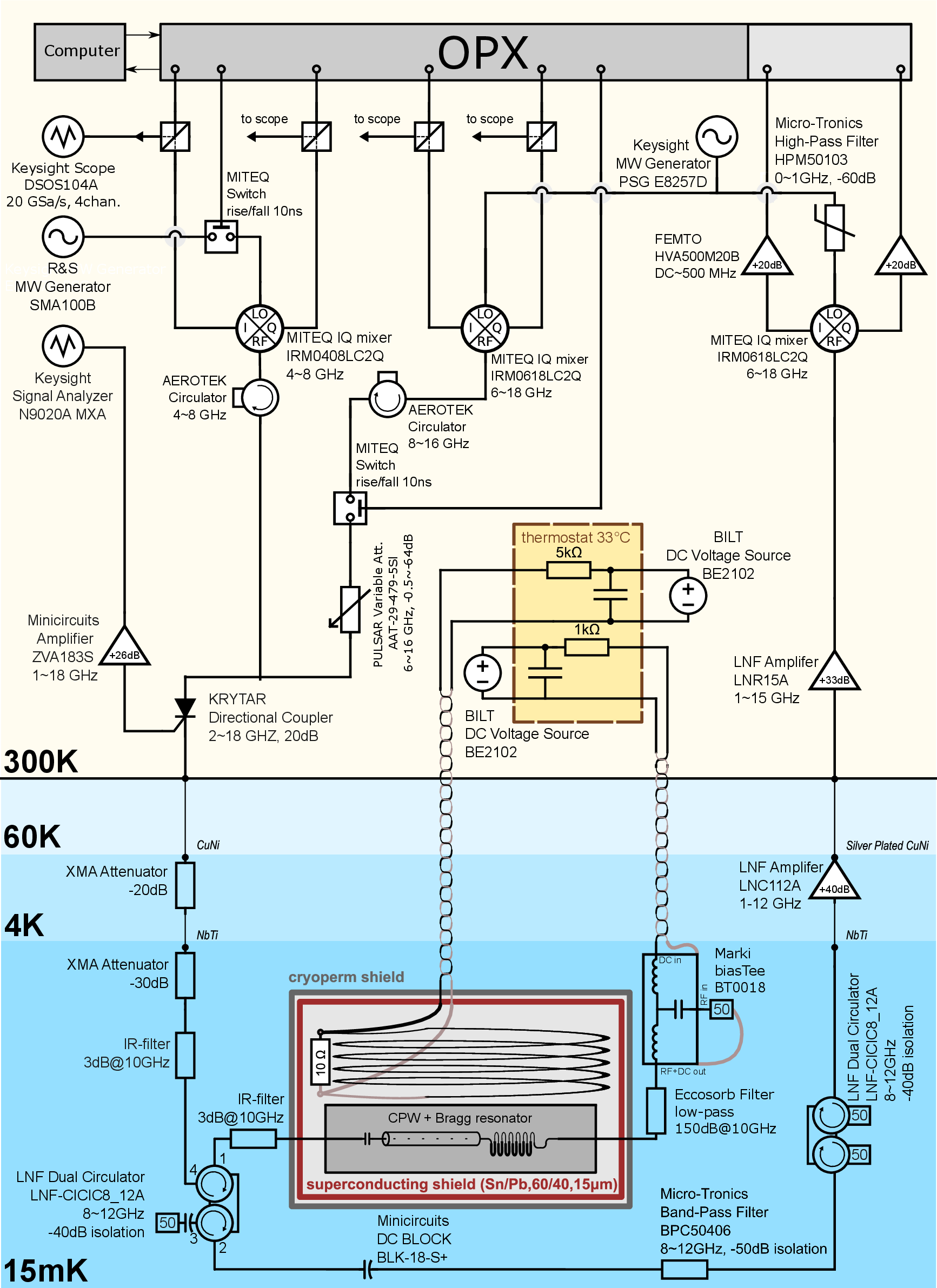}\caption{Experimental Setup\label{extfig1}}
\end{figure*}

\begin{figure*}[p]
\includegraphics[width=\textwidth]{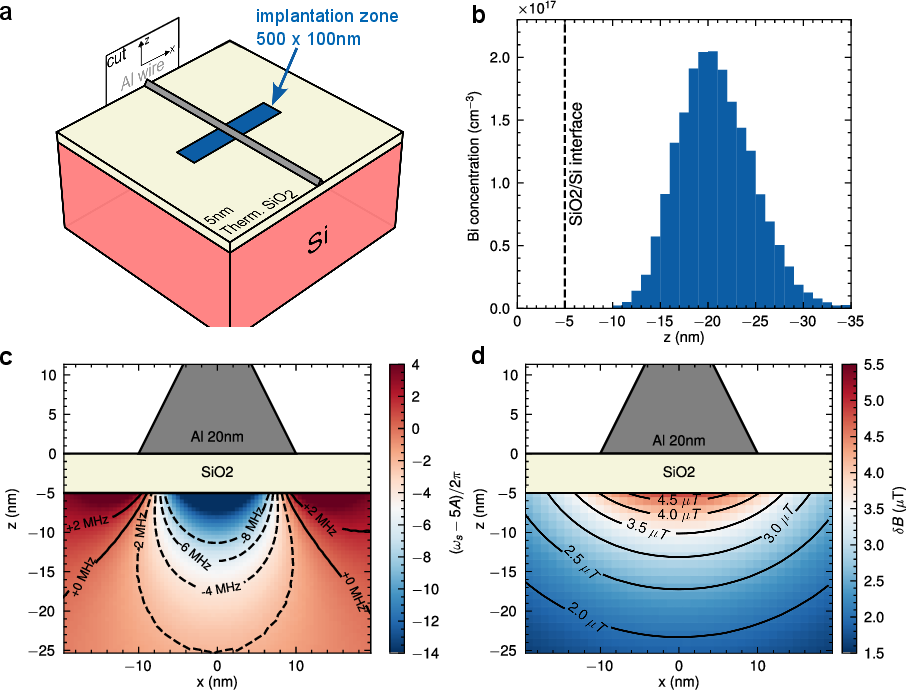}\caption{\textbf{ a}, Top view of the implantation
region crossed by a thin aluminum wire. \textbf{b}, The bismuth implantation
profile calculated using Monte Carlo simulation (SRIM) for an ion
energy of 26 keV into a silicon substrate covered with $\SI 5{\nm}$
of thermally grown $\protect\siox$. \textbf{c}, Strain-induced shift
of the ESR absorption due to the presence of a $\SI{20}{\nm}$
triangle wire using similar assumptions than Ref. \citep{PhysRevApplied.9.044014}
and calculated using Comsol Multiphysics software. \textbf{d}, Cut
of a $\SI{20}{\nm}$ triangle wire and simulated map of the
magnetic fluctuations calculated assuming current fluctuations $\delta I=\SI{300}{\nano\ampere}$
calculated using Comsol Multiphysics software. \label{extfig2}}
\end{figure*}

\begin{figure*}[p]
\includegraphics[width=\textwidth]{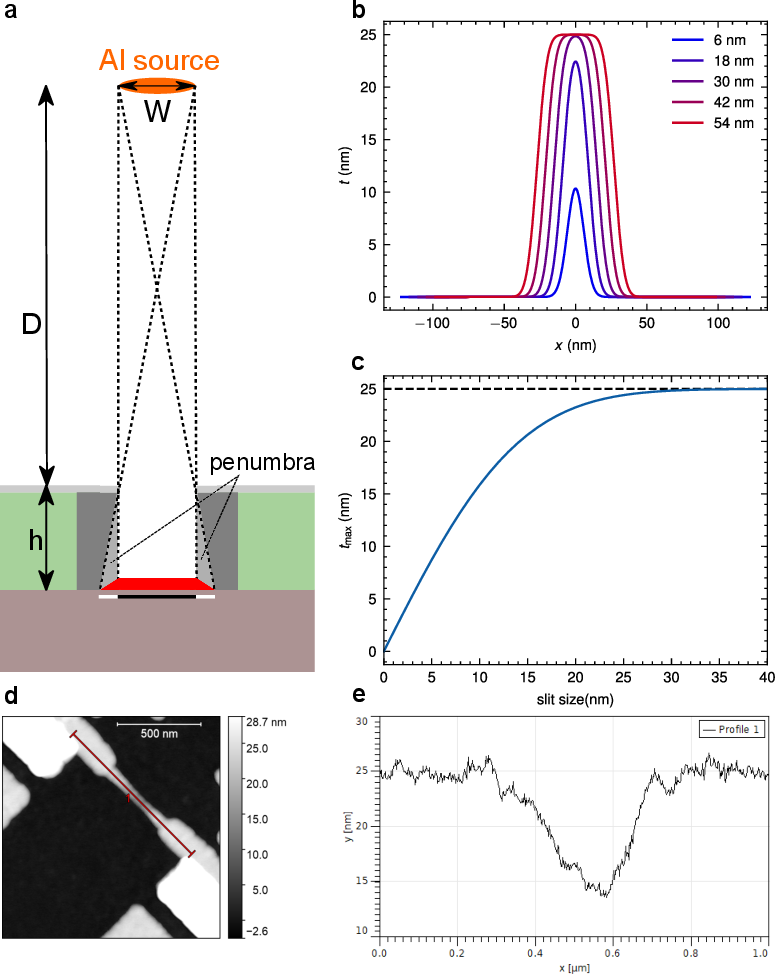}\caption{\textbf{a}, Scheme of the penumbra
formation. In our experimental system, the evaporated metal is coming
from a source with a gaussian distribution of diameter $W=\SI{10}{\milli\meter}$
at a distance $D=\SI{500}{\milli\meter}$ from the sample
and a mask suspended at height $h=\SI{650}{\nano\meter}$,
giving constrictions with minimal width $h\frac{W}{D}=\SI{13}{\nano\meter}$.
\textbf{b}, Evaporated profile for different slit size assuming a
quantity of $\SI{25}{\nano\meter}$ of evaporated aluminum.
\textbf{c}, Maximum constriction thickness as a function of slit size.
\textbf{d}, AFM micrograph of a test constriction. \textbf{e}, Profile
along the cut 1 in \textbf{d} \label{extfig3}}
\end{figure*}

\begin{figure*}[p]
\includegraphics[width=\textwidth]{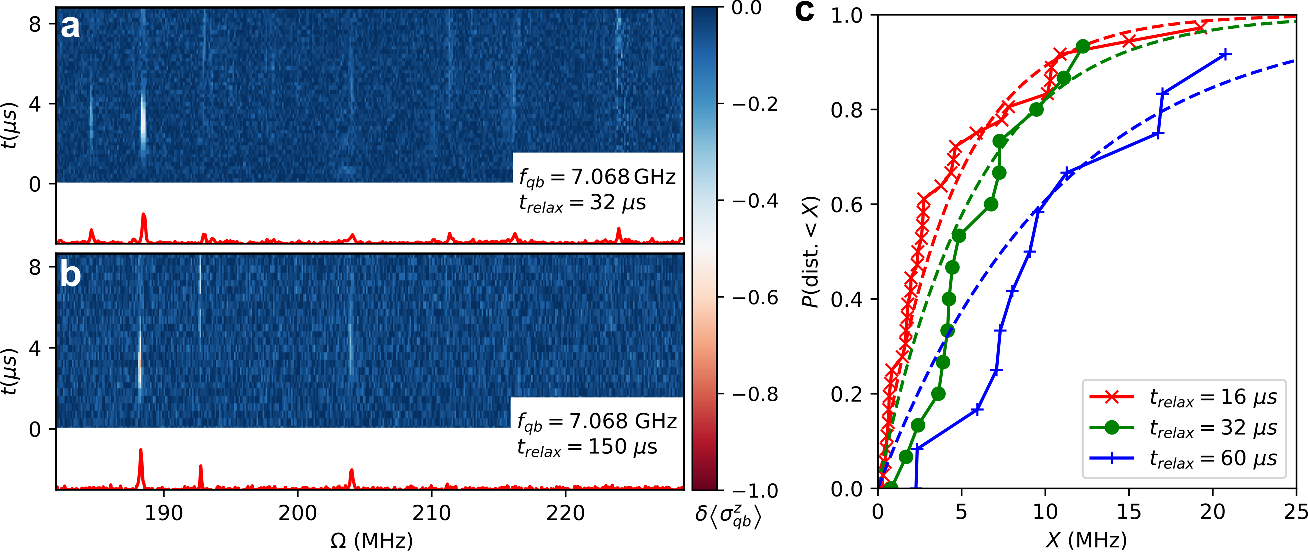}\caption{
\textbf{a},
Detection of two-level systems (TLS) using protocol shown in Fig. \ref{fig3}a with  a flux qubit at $f_{qb}=\SI{7.068}{\giga\hertz}$ and $t_\text{relax}=\SI{32}{\micro\second}$. 
\textbf{b},
Detection of two-level systems using protocol shown in Fig. \ref{fig3}a with  a flux qubit at $f_{qb}=\SI{7.068}{\giga\hertz}$ and $t_\text{relax}=\SI{150}{\micro\second}$. Only TLSs with relaxation time longer than $\SI{150}{\micro\second}$ are still visible. 
\textbf{c},
Probability $P(d<X)$ for detecting the nearest TLS at a distance d smaller than X for different values of $t_\text{relax}$. The dashed lines correspond to uniform distribution densities $P(d<X)=1-\exp(-\rho X)$ with $\rho=\SI{0.22}{\mega\hertz}^{-1}$ (in red), $\rho=\SI{0.17}{\mega\hertz}^{-1}$ (in green) and $\rho=\SI{0.09}{\mega\hertz}^{-1}$ (in blue).
\label{extfig4}
}
\end{figure*}

\begin{figure*}[p]
\includegraphics[width=\textwidth]{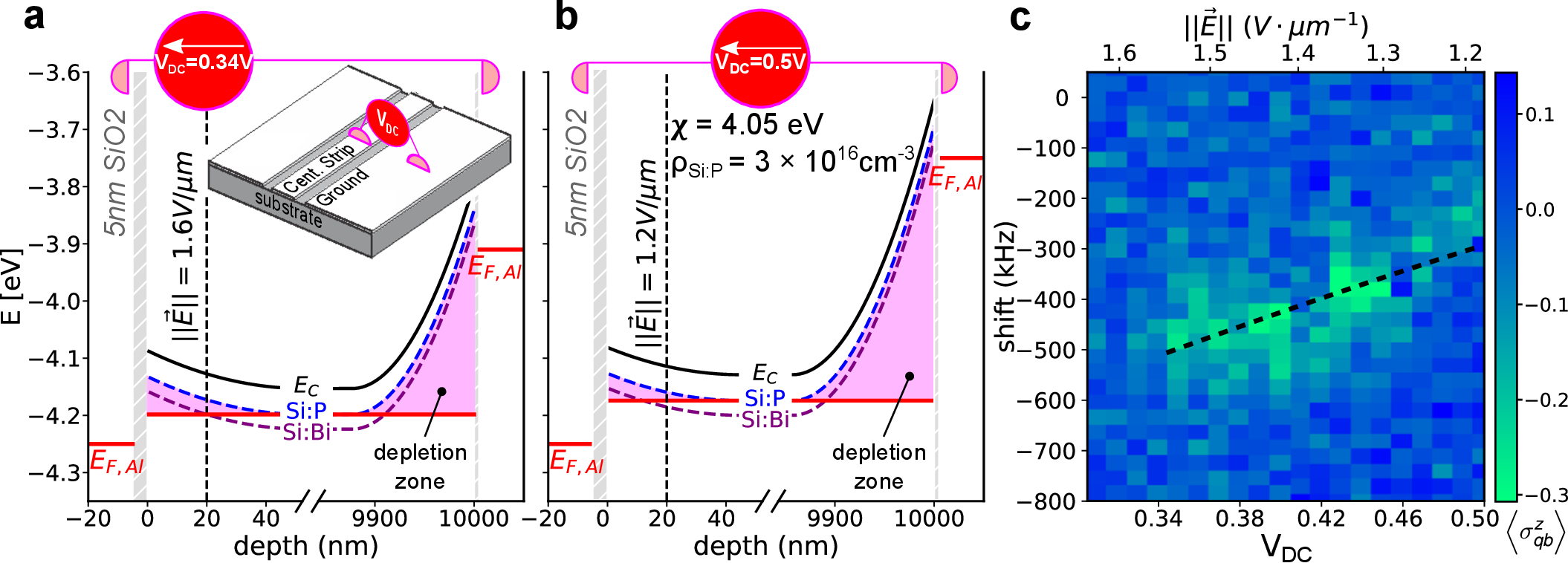}\caption{\textbf{a}, 1D numerical simulation of the depletion region using Poisson equation at $V_{DC}=\SI{0.34}{\volt}$ assuming that the bismuth donor situated at a depth of $\SI{20}{\nano\meter}$ below the surface is ionized at that voltage. The electric field seen by the donor is calculated to be  $\|E\|=\SI{1.6}{\volt\per\micro\meter}$. \textbf{b}, 1D numerical simulation of the depletion region using Poisson equation at the working point $V_{DC}=\SI{0.5}{\volt}$. The electric field seen by the donor is calculated to be  $\|E\|=\SI{1.2}{\volt\per\micro\meter}$. \textbf{c}, Experimental shift of the spin spectral line versus $V_{DC}$. The dashed line corresponds to the stark shift expected for a bismuth donor according to Ref. \citep{PhysRevB.90.195204}. 
\label{extfig5}}
\end{figure*}

\backmatter
\bmhead{Acknowledgements}
This research was supported by the Israeli Science Foundation under
grant numbers 426/15, 898/19 and 963/19. We acknowledge T. Schenkel
for kindly providing us with the isotopically purified Isonics wafer
used in this work. M. Stern wishes to thank P. Bertet for fruitful
discussions all over the years and A. Morello for his good advice
``gates are your friends !''.


\end{document}